\RequirePackage{fix-cm}

\documentclass[smallextended]{svjour3}

\smartqed  % flush right qed marks, e.g. at end of proof

\usepackage{graphicx}
\graphicspath{{figures/}}
\usepackage{wrapfig}
\usepackage{float}
\usepackage{caption}
\usepackage{hyperref}
\usepackage{tabularx}
\usepackage{array}
\usepackage{amsmath}
\usepackage{amssymb}
\usepackage{color}
\usepackage{mathtools}
\usepackage{slashed}
\usepackage{pdfpages}
\usepackage{listings}
\usepackage{siunitx}
\usepackage{latexsym}
\usepackage{multirow}
\usepackage{lineno}

\begin{document}

\title{Study of Particle Multiplicity of Cosmic Ray Events using 2\,m\,$\times$\,2\,m Resistive Plate Chamber Stack at IICHEP-Madurai}

\author{Suryanarayan Mondal \and
  V. M. Datar \and
  Gobinda Majumder \and
  N. K. Mondal \and
  S. Pethuraj \and
  K. C. Ravindran \and
  B. Satyanarayana
}

\institute{
  Suryanarayan Mondal \and V. M. Datar \and Gobinda Majumder \and S. Pethuraj \and K. C. Ravindran \and B. Satyanarayana \at
  Tata Institute of Fundamental Research, Dr. Homi Bhabha Road, Mumbai, India \\
  \email{suryamondal@gmail.com, vivek.datar@gmail.com, majumder.gobinda@gmail.com, spethuraj135@gmail.com, ravitifr@gmail.com, bsn@tifr.res.in}
  \and
  Suryanarayan Mondal \and S. Pethuraj \at
  Homi Bhaba National Institute, Anushaktinagar, Mumbai, India \\
  \and
  N. K. Mondal \at
  Saha Institute of Nuclear Physics, Bidhannagar, Kolkata, India\\
  \email{nabak.mondal@gmail.com}
}

\date{Received: date / Accepted: date}
\maketitle

\begin{abstract}
  An experimental setup consisting of 12 layers of glass Resistive Plate Chambers 
  (RPCs) of size 2\,m\,$\times$\,2\,m has been built at IICHEP-Madurai
  (\ang{9;56;14.5}\,N \ang{78;00;47.9}\,E) to study the 
  long term performance and stability of RPCs produced on a large scale in Indian 
  industry. This setup has been collecting data triggered by the passage of charged 
  particles. The measurement of the multiplicity of charged particles due to cosmic ray 
  interactions are presented here. Finally, the results are compared with 
  different hadronic models of the CORSIKA simulation.
\end{abstract}

\section{Introduction}

The 50 kton INO-ICAL\cite{inowhite} is a proposed underground high energy physics
experiment at Theni, India (\ang{9;57;50.1}\,N \ang{77;16;21.8}\,E) to study the
neutrino oscillation parameters using atmospheric neutrinos. The primary aim of
the experiment is to determine the sign of the mass-squared difference
\mbox{$\Delta m^2_{32}$ $\left(=m^2_3-m^2_2\right)$} using matter effects. The ICAL
 detector can also be used to probe the value of \mbox{leptonic CP-phase
$\left(\delta_{cp}\right)$} and last but not the least to search for physics beyond
 the standard model using neutrino oscillations. The Resistive Plate Chamber
(RPC)\cite{rpc_p1,rpc_p2} has been chosen as the active detector element for the
ICAL detector. As part of the ICAL R\&D program, a 12-layer stack of
2\,m\,$\times$\,2\,m RPCs have been operational at IICHEP, Madurai
(\ang{9;56;14.5}\,N \ang{78;00;47.9}\,E, on the surface, 160\,m above mean sea level)
since the last few years. The various detector properties like position and time
resolution of RPCs, detector efficiencies, strip multiplicities, detector noise,
etc. are studied using this RPC stack to understand the performance and long term
stability of the RPCs. The same data are also used to study the cosmic ray
muons \cite{pethu1}. The data collected near magnetic equator gives us vital
information regarding the capabilities of the simulation packages.

High energy primary cosmic rays originating in outer space continuously interact 
with the earth's atmosphere. These cosmic rays consist of mostly protons with a 
smaller fraction of higher \mbox{Z-Nuclei} elements \cite{pdgspectra1}. The angular 
distribution of primary cosmic rays is more or less isotropic at the top of the 
earth atmosphere. The energy spectrum of the primary cosmic rays follows a 
power-law spectrum, $dN/dE \propto E^{-\gamma}$, where power-law parameter, 
$\gamma \sim $ 2.7. The shower of particles (called secondaries) consists mainly 
of \mbox{pions $\left(\pi^{\pm}/\pi^0\right)$} and 
\mbox{kaons $\left(K^{\pm}\right)$} which
are produced due the interactions of primary cosmic rays with atmospheric nuclei. 
The neutral pions mainly 
decay via electro-magnetic interactions, $\pi^0 \rightarrow \gamma+\gamma$ whereas 
the charged pions decay to muons and neutrinos via weak-interactions, 
$\pi^+ \rightarrow \mu^+ + \nu_{\mu}$ and 
$\pi^- \rightarrow \mu^- + \bar{\nu}_{\mu}$. The kaons also decay to muons and 
neutrinos and to pions in different branching fractions. Most of the pions and 
kaons decay in flight and do not reach the earth's surface, whereas only a small
fraction of resultant muons decay into electrons and neutrinos, 
$\mu^+ \rightarrow e^+ + \nu_{e} + \bar{\nu}_{\mu}$ and 
$\mu^- \rightarrow e^- + \bar{\nu}_{e} + \nu_{\mu}$. The $\gamma$, $e^{\pm}$ do not
reach the detector directly as they interact with the roof of the laboratory and 
create electromagnetic showers. Thus, muons are the most abundant charged particle 
from cosmic ray showers detected in the present setup. These atmospheric muons are 
produced at high altitude (average height of 20\,km) in the atmosphere and lose 
almost 2\,GeV energy via ionisation loss in the air before reaching the ground. The 
density of charged particles (mainly muons) per unit surface area at the earth's 
surface depends on the composition of primary cosmic ray, power-law parameter
($\gamma$) as well as the model of hadronic interactions at high energy which is
not accessible in the laboratory.

The principal aim of this work is to observe the charged-particle multiplicity in
the atmospheric muon data collected at IICHEP, Madurai and compare it with the air
shower simulation.
Most of the secondary particles reaching the surface of the earth are having low
energies which are produced by the interactions of the cosmic primaries with the
air at center-of-mass energies much lower than 10\,GeV, where
the hadronic interaction is better understood \cite{corsika763}.
The distribution of charged-particle multiplicity is thus
expected to be dominated by the composition of primary cosmic ray and the spectral
index, $\gamma$.

In this paper, the detector setup is described in Section~\ref{sec:detector}. 
The Monte-Carlo simulation used to study the multiplicity is explained in 
Section~\ref{sec:montecarlo}, where primary cosmic ray interactions are simulated
using the CORSIKA Package\cite{corsika763} and interactions of the particle with
detector material is simulated using the GEANT4 toolkit\cite{geant4}.
The method of Hough Transformation which is used to find the trajectory of charged 
particle and the data selection criteria used for the analysis are discussed in 
Section~\ref{sec:reconstrction}. Finally, the results of the analysis are 
summarised in Section~\ref{sec:result}.

\section{Detector Setup} \label{sec:detector}
\begin{figure}[h]
  \centering
  \includegraphics[width=0.99\linewidth]{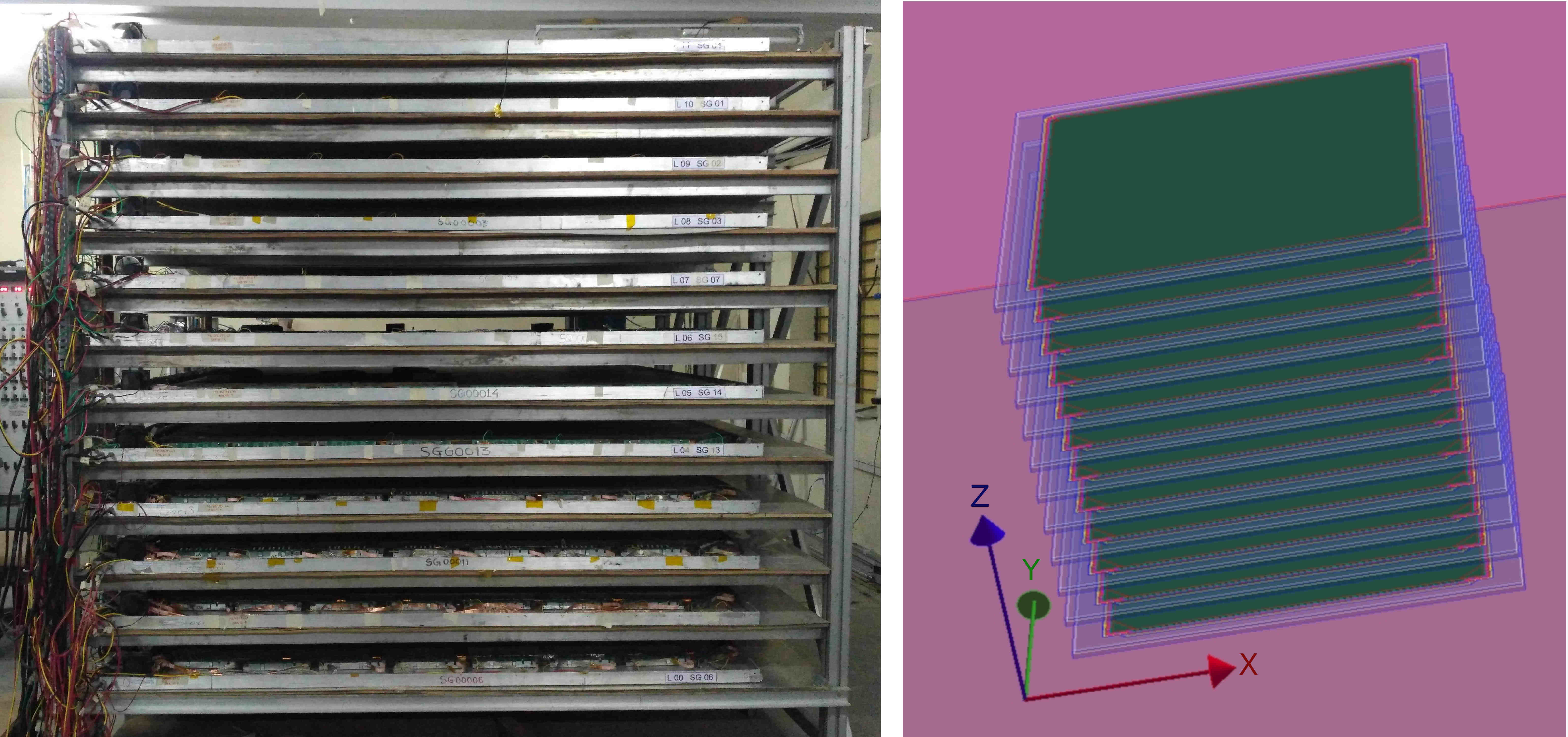} 
  \caption{The detector stack with 12 layers of RPCs,
    (left) experimental setup and (right) Geant4 detector geometry of stack.}
  \label{fig:stack}
\end{figure}

The RPC stack operational at IICHEP, Madurai consisting of 12 RPCs stacked 
horizontally with an inter-layer gap of 16\,cm is shown in Figure~\ref{fig:stack}.
An RPC gap is made of two glass electrodes of thickness 3\,mm with a gap of 2\,mm 
between them. This gap is maintained using 2mm thick poly-carbonate buttons. The 
glass gap is sealed on the outer edge to make it air-tight. A non-flammable mixture
of gas is continuously flown inside the glass gaps which serve as the active 
medium of the detector. In avalanche mode, the mixture of gas consists of 
R134a (95.2\%), iso-C$_4$H$_{10}$ (4.2\%) and SF$_6$ (0.3\%). Both the outer 
surfaces of the glass gap are coated with a thin layer of graphite. The RPCs are 
operated by applying a differential supply of $\pm$\,5\,kV to achieve the desired 
electric field. The avalanche created by the ionisation energy loss of charged 
particles in the RPCs induces signals in the two orthogonal pickup panels placed on 
both sides of the glass gaps labelled as X-side and Y-side. The pickup panels are 
made of parallel copper strips of width 28\,mm with 2\,mm gap between two 
consecutive strips. The RPCs used in this detector stack are of the size of 
1790\,mm\,$\times$\,1890\,mm. There are 60 strips on the X-side and 63 strips on 
the Y-side for each layer.

\begin{figure}[h]
  \includegraphics[width=1.0\linewidth]{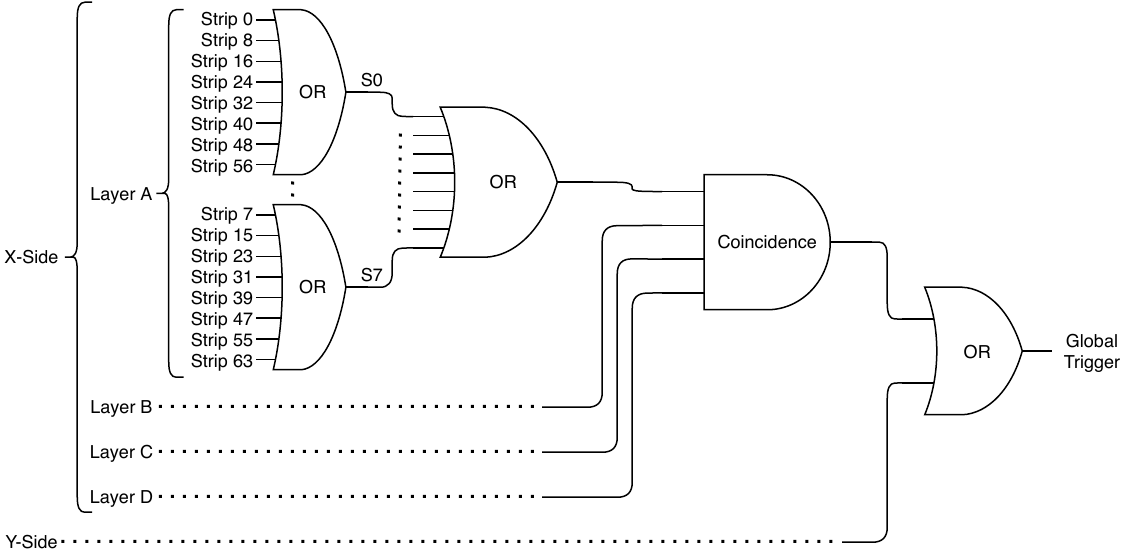} 
  \caption{Trigger Scheme of the Detector Setup.}
  \label{fig:trigger}
\end{figure}

The induced signals from the pickup strips are amplified and discriminated by a 
charge sensitive NINO\cite{nino} amplifier-discriminator board. In layer 
11 (top-most layer), the ANUSPARSH ASIC\cite{anusp} which is a CMOS, 8-channel,
high speed, low power amplifier-discriminator designed for avalanche mode of
operation for RPCs is used to study its performance.
The discriminated signals from these boards are
passed to the FPGA-based RPCDAQ-board. The individual signals from every 
8$^{th}$ strips are \emph{OR}ed to get pre-trigger signals (S0 to S7), which
are passed to the Trigger system module via Signal Router Board.
The coincidence is done for X- and Y- sides independently in the Trigger Logic Boards.
The Global Trigger is then generated by the Global Trigger Logic Board (GTLB)
by \textit{OR}ing the coincidences formed in X- and Y-side.
The trigger scheme of the detector setup contributed by 4 RPC layers is
elastrated in the Figure~\ref{fig:trigger}.

\begin{figure}[h]
  \includegraphics[width=1.0\linewidth]{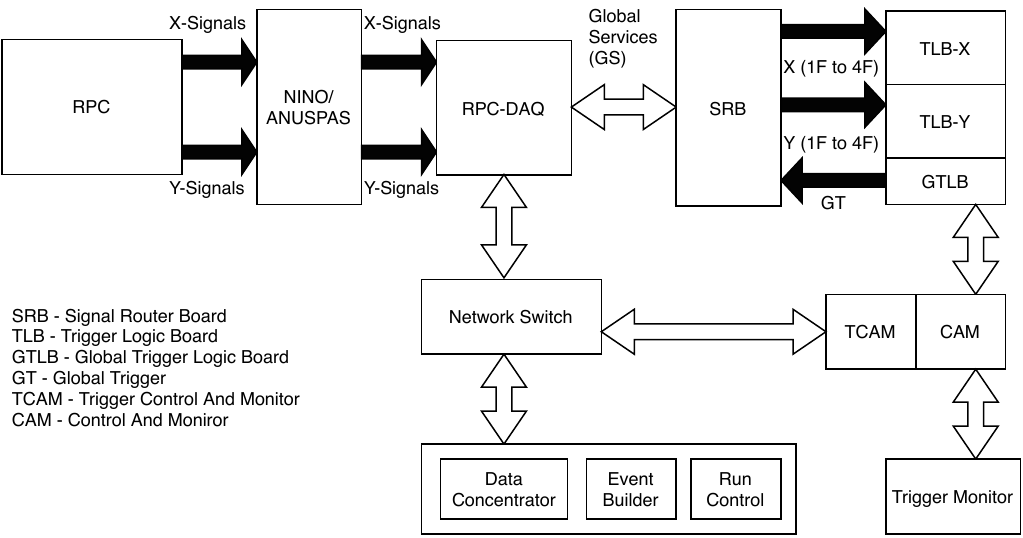} 
  \caption{The Flow of Signals in the Detector Setup.}
  \label{fig:sigflow}
\end{figure}

The hit signals in the RPCDAQ board stretched to
1\,$\mu$s to overcome trigger latency of 770\,ns from Trigger System to RPCDAQ.
Based on the arrival of trigger signals to RPCDAQ, the event signals are latched
and sent to the Data Concentrator and the Event Builder via Network Switch. The flow of
signals in the Detector setup are shown in Figure~\ref{fig:sigflow}.
The detailed description of signal processing and the Data Acquisition system 
(DAQ) can be found in \cite{elec1}. The present work is based on the coincidence 
of pre-Triggered signals from layers 4, 5, 6 and 7 within a coincidence window of
100\,ns. While the layers in the middle portion of the detector are used to form the
trigger, the layers in both the top and bottom portions of the detector also
contributes in forming an event, which in turns maximises the length of the track in
the detector.

Although the coincidence window is 100\,ns, the signals from any other particles as
well as noise signals occurring within a time window of -230\,ns to 770\,ns after
generation of the trigger are also recorded due to stretching of the hit
signals and the trigger latency. An event typically contains
hit information (one logic bit per strip indicating the signal in that strip) for
each strip and 16 time signals for each layer; 8 for X-side and other 8 for Y-side.
One TDC (Time-to-Digital~Converter) channel with least-count of 0.1\,ns
records time signals coming from every alternating 8$^{th}$ strips (S0 to S7).
Approximately 250 millions of cosmic ray events recorded in the detector during the
total observation period between August 23, 2017 to September 8, 2017 with a trigger
rate of $\sim$230\,Hz are used for the analysis.

\section{Monte-Carlo Simulation} \label{sec:montecarlo}
The Monte-Carlo Simulation for this study has been performed in two stages. 
Extensive Air Shower (EAS) has been simulated by the CORSIKA. 
The information of daughter particles generated by EAS at the earth's surface
level has 
been extracted and used as the input to the detector simulation. The detector 
simulation has been performed using the GEANT4 toolkit. 

\subsection{Extensive Air Shower}
The CORSIKA (COsmic Ray SImulations for KAscade) \cite{corsika763} developed to
study the 
evolution of EAS in the atmosphere initiated by cosmic ray particles. Though the 
CORSIKA has been developed for a specific experiment, it has now developed into a 
tool that is used by many groups studying cosmic rays and EAS. In the present 
scenario, existing extrapolation of hadronic interaction models of high energy 
particles in EAS is based on various theoretical models, which has large 
uncertainties. The current experimental data at the collider experiment is 
insufficient to predict the hadronic interactions at very high energies. In the 
CORSIKA package, the several different hadronic interaction models are available. 
In this study, for simulating the behaviour of hadrons for higher energy range, the 
QGSJET (Quark Gluon String model with JETs)\cite{corsika763} has been adopted and 
for the low energy range (less than 80\,GeV in laboratory frame), the GHEISHA model 
has been used. 

In this study, the primary cosmic ray shower has been simulated using the 
CORSIKA(v7.6300) Package. The energy of the primary rays in the CORSIKA is
generated using the power-law spectrum, $E^{-2.7}$, within the energy range of 
\mbox{$10$--$10^{6}$\,GeV} for different primaries (H, He, C, O, Si and Fe). The 
zenith and azimuth angle of primary particles are generated uniformly within the 
range of \mbox{$0$--$85^\circ$} and \mbox{$0$--$360^\circ$}, respectively. The 
magnetic rigidity cutoff has been implemented according to the location of the 
detector.
The minimum energy cutoff for hadrons, muons, electrons and photons in the 
simulation is kept at 50\,MeV, 10\,MeV, 1\,MeV and 1\,MeV, respectively. These 
cutoff values are much smaller than the minimum momentum cutoff for the charged
particles in the vertical direction, \mbox{$\sim 110$\,MeV} at 1\,GeV energy,
which is mainly due to 22\,cm of 
concrete roof of the building where the detector is placed \cite{pdgspectra1}.

\begin{figure}[h]
  \centering
  \includegraphics[width=0.7\linewidth]{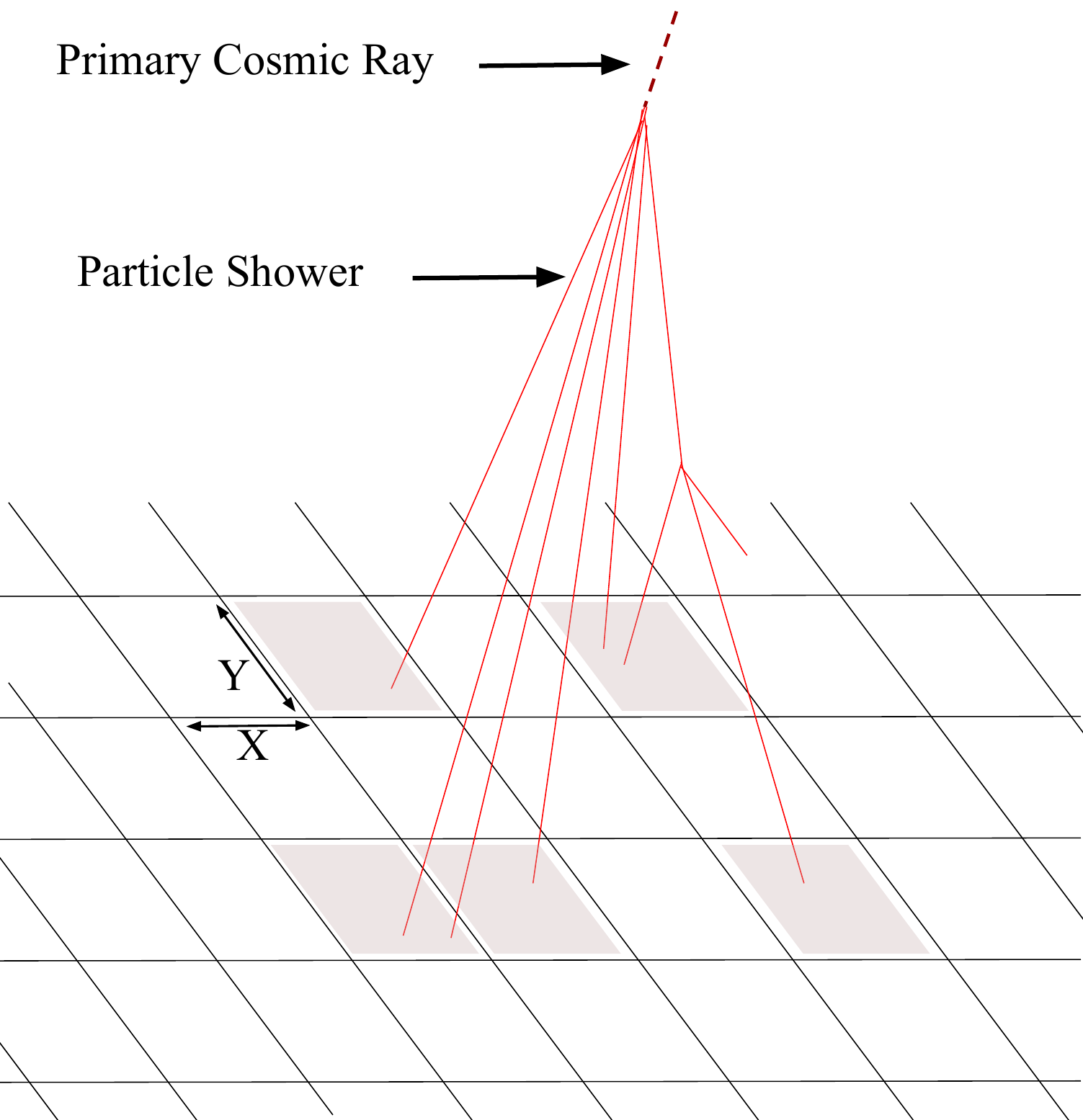} 
  \caption{Shower of particles initiated by primary cosmic ray reaching observation surface.}
  \label{fig:eas}
\end{figure}

The particles generated by the CORSIKA at the observation surface are provided
as an input to the detector simulation. The observation plane has been divided
into squares of the size of 2\,m\,$\times$\,2\,m which is shown in Figure~\ref{fig:eas}.
An event is formed using the information of the particle(s) passing through each
of these rectangles shown as shaded regions in Figure~\ref{fig:eas}.

\subsection{Detector Simulation}
The detector simulation has been performed using the GEANT4(v4-10.0.2) toolkit. The 
events from the CORSIKA simulation are propagated in the detector simulation. A 
realistic depiction of the detector setup including the building where the detector 
is housed has been constructed in the GEANT4 environment. The properties of various 
materials of the detector components and the laboratory building are chosen based 
on the knowledge of the setup. The uncertainty of the material budget is taken as a 
systematic error. The standard physics processes of matter-particle interactions 
like electromagnetic, ionisation, decay and hadronic interactions
(QGSP$\_$BERT$\_$HP), which are available within the GEANT4 toolkit \cite{geant4}
are implemented in the simulation.

The various detector's parameters (efficiency, noise, strip multiplicity and 
resolution) are calculated with the help of cosmic ray passing through the detector.
The current setup is a tracker type detector consisting of 12 layers of RPCs.
When one layer of the setup is studied for the detector parameters;
the rest of the layers in the setup serve as the tracker for the passing particles.
Since the layers forming the coincidence cannot be studied for the detector
parameters, the data sets with two more different trigger combinations (using
layers 1,2,3,4 and 7,8,9,10 to form coincidences) of the duration of one day,
are used in order to study all the RPCs in the setup.

\begin{figure}[h]
  \centering
  \includegraphics[width=1.\linewidth]{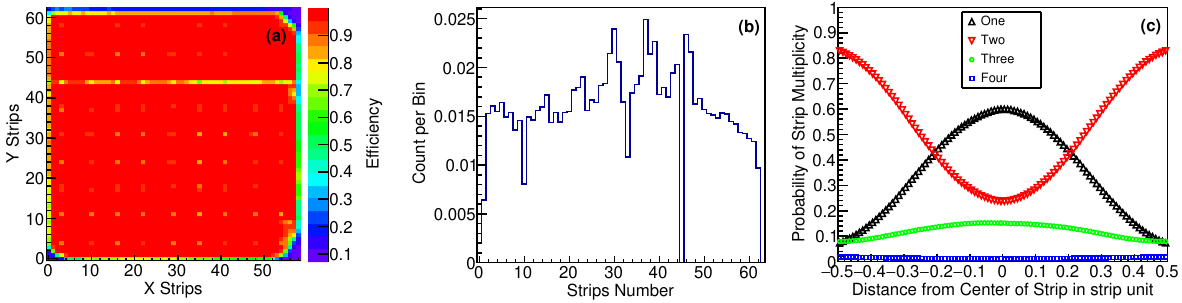} 
  \caption{(a) Efficiency, (b) Noise and (c) Multiplicity profile of Y-side of 
               Layer-2 RPC gap.}
  \label{fig:layer2y}
\end{figure}

Each of the RPC detectors can be divided into a matrix of pixels of size
3\,cm$\times$\,3\,cm. The efficiencies of each of the pixels are calculated and
represented as the efficiency map for each of the RPCs. The track(s) of the
particle(s) in an event is fitted with the straight line by excluding the RPC layer
being studied from the fit. For each pixel in that RPC layer, the total number of
particles passed through it is estimated from the extrapolated position in that
layer. The ratio of the number of events with valid hit signal in X- (or Y-) strips
of that pixel to the total number of particles passed through it is taken as the
efficiency for X- (or Y-) side of that pixel.
The noise in a RPC is defined as the hits occurred farther
from the expected position of the passing particle. The strip
multiplicity profile is defined as the probability of sharing signal
between neighbouring strips with respect to the hit position from the
centre of the strip. The strip multiplicity is discussed in the next section.
A detailed study of these parameters is presented in \cite{pethu1}.
The efficiency map, noise and strip multiplicity profile for one of the RPCs
in the stack is shown in Figure~\ref{fig:layer2y}.
These observed detector parameters are included in the digitisation stage of the
detector simulation. The events from the cosmic ray data and the detector
simulation are reconstructed using an algorithm based on Hough Transformation
which is discussed in the next section.

\section{Event Reconstruction and Data Selection}\label{sec:reconstrction}
For the event reconstruction, the strips hits are analysed separately, in the 
2-dimensional projections namely, \mbox{X--Z} and \mbox{Y--Z} side. When a charged 
particle passes through the RPC, the number of strips on which signal is induced 
depends on the gain of the gas gap. The sharing of the induced signal between the 
neighbouring strips is the main reason for the observed strip multiplicity shown in 
Figure~\ref{fig:layer2y}(c). During the study, the position resolution is 
calculated for different strip multiplicities of 1, 2, 3, and 4 and the values 
observed are $\sim$6\,mm, $\sim$8\,mm, $\sim$12\,mm and $\sim$22\,mm respectively. 
The position resolution for strip multiplicity more than four is larger than the 
pitch of the strip (3\,cm). In the present study, the clusters are formed with a 
maximum of 4 consecutive strips as the position resolution for higher multiplicities
is found to be worse. A layer which has more than 15 strip hits or more than 10 
clusters are tagged as `noisy layer' and not considered in track reconstruction.
The first criterion has been chosen near the maximum number of possible hits 
if 4 tracks pass through a RPC. In fact, the maximum number of tracks reconstructed
in an event is 4 which is discussed in the Result section. The second cut is set at
the first cut divided by the average strip multiplicity $\left(\sim 1.5\right)$
in the detector stack to reject noisy events passed through the first cut.
An event which has more than 3 noisy layers is considered as `noisy event' and 
discarded. This cut is set at 3 layers which is 25\% of maximum layers available
for the event reconstruction. This cut has been set by balancing the performance
of the reconstruction method and number of events lost due to this cut.

\begin{figure}[h]
  \includegraphics[width=1.0\linewidth]{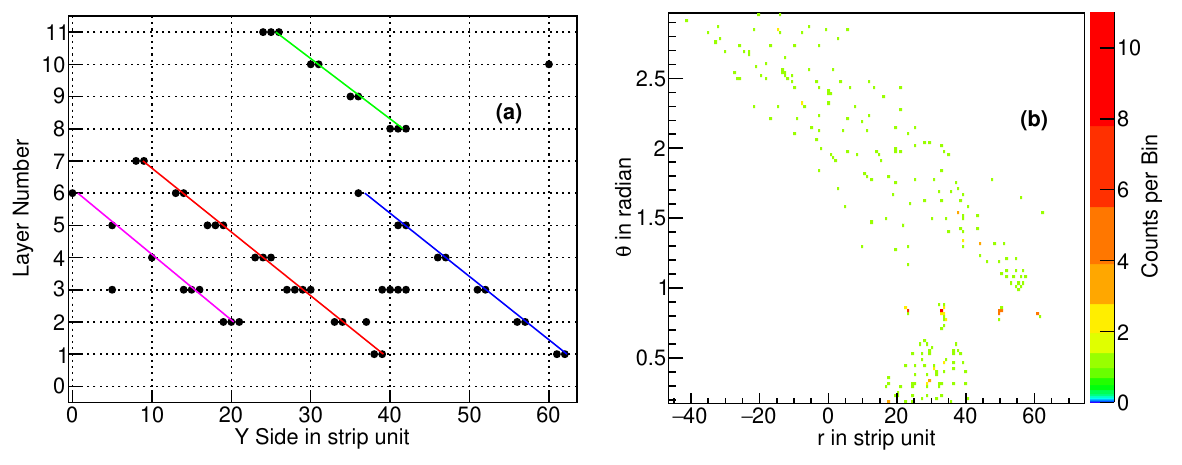} 
  \caption{(a) Projection of an event in the detector and (b) populated 
               $r$-$\theta$~plane using this event.}
  \label{fig:houghPl}
\end{figure}

In the first step of track reconstruction, the clusters associated with different 
tracks are grouped using the method of Hough Transformation\cite{hought1,hought}. The 
equation of the straight line, used to find the association between the hits, is 
given as,
\begin{equation}
  r=z\cos\theta+x\left(/y\right)\sin\theta. \label{eq:hough}
\end{equation}
The \mbox{$r$-$\theta$} plane (called as Hough Space) is populated using the 
concept of Cellular Automaton\cite{cellular}. For a sample event shown in 
Figure~\ref{fig:houghPl}(a), the populated \mbox{$r$-$\theta$} plane is presented 
in Figure~\ref{fig:houghPl}(b). 
The advantage of using Cellular Automaton technique is the significant reduction
of computation time to find a trajectory in the event. This method can detect all 
the tracks avoiding the noise hits as shown in Figure~\ref{fig:houghPl}(a).

\begin{figure}[h]
  \includegraphics[width=1.0\linewidth]{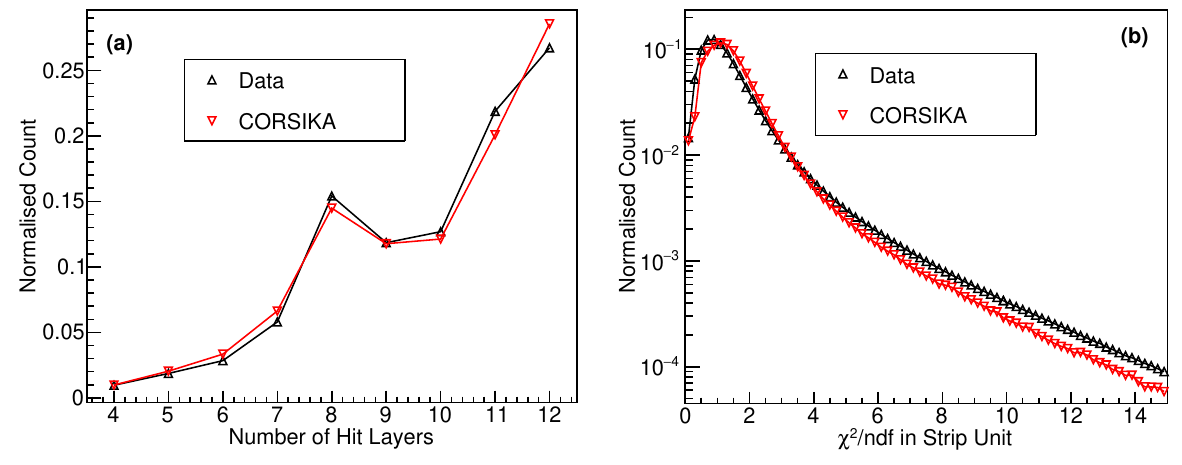} 
  \caption{(a) Number of hit layer and (b) $\chi^2$/ndf of straight line fit.}
  \label{fig:chi2ndf}
\end{figure}

\begin{figure}[h]
  \includegraphics[width=1.0\linewidth]{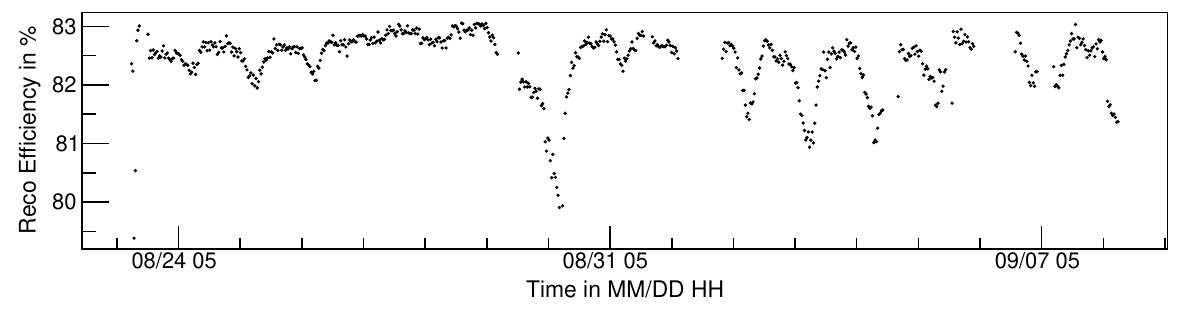} 
  \caption{Variation of reconstruction efficiency of the detector with time.}
  \label{fig:stackineffi}
\end{figure}

The tracks are identified using the Hough Transformation are fitted by a straight line 
given by the equation,
\begin{equation}
  x\left(/y\right)=mz+c \label{eq:plain}
\end{equation}
where $m$ and $c$ are the slope and the intersect, respectively.
The number of detector layers in the fit and $\chi^{2}$/ndf of the fit are shown in 
Figure~\ref{fig:chi2ndf}(a) and \ref{fig:chi2ndf}(b) respectively.
A track is considered as reconstructed if the $\chi^{2}$/ndf is less than 10 and 
there are more than 4 layers in the track. The reconstruction efficiency is defined
as the ratio of the number of events with at-least one reconstructed track with the 
total number of triggered events. The reconstruction efficiency as a function of 
time is shown in Figure~\ref{fig:stackineffi}.
It can be observed that the reconstruction efficiency varies periodically which 
is correlated to the variation of atmospheric pressure and temperature
\cite{rpcleak}. This periodic change in 
efficiency does not affect the relative ratio of multiple track events. The pure 
multiple track events are $\sim$0.01\% of triggered events. Out of the total
triggered events, 6--7\,\% of events are due to noise and hadronic showers
initiated at the roof. Any such ambiguous events are rejected by the selection
criteria discussed in the following.

\begin{figure}[h]
  \includegraphics[width=1.0\linewidth]{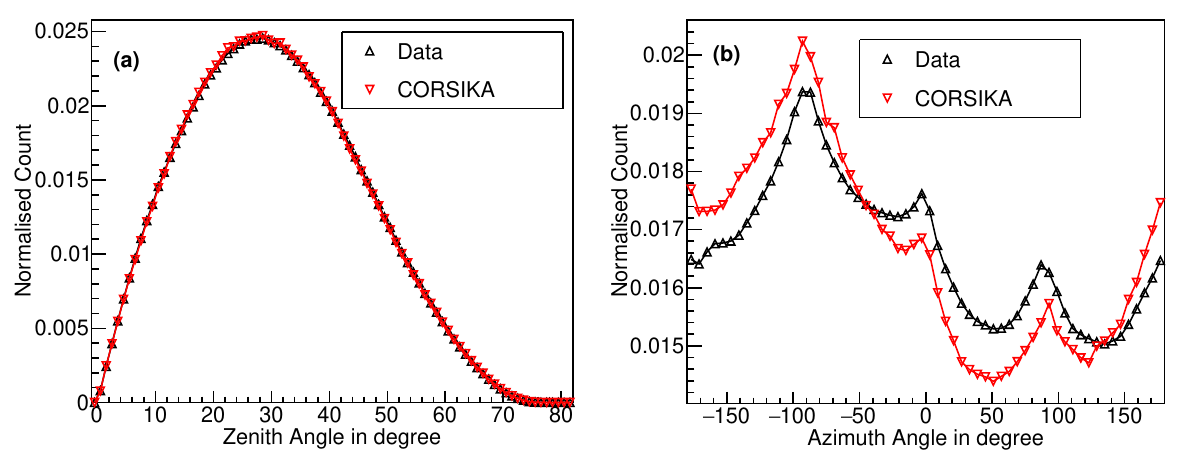} 
  \caption{(a) Zenith and (b) Azimuth Angle of cosmic rays reaching the detector 
stack.}
  \label{fig:thetaphi}
\end{figure}

The zenith and azimuth angle distributions of the reconstructed tracks are 
presented in Figure \ref{fig:thetaphi}(a) and \ref{fig:thetaphi}(b), respectively.
The projections from both X--Z and Y--Z sides are combined to produce final 
3-dimensional track(s).
The `Fake Tracks' may arise while combining the projections. A Fake track is defined
as a reconstructed track which does not correlate to the X- and Y-side projection of
the trajectory of a true particle but instead is formed due to the mismatch of the
projections from separate particle. Any such `Fake Tracks' are discarded by using
the timing information.

\begin{figure}[h]
  \includegraphics[width=1.0\linewidth]{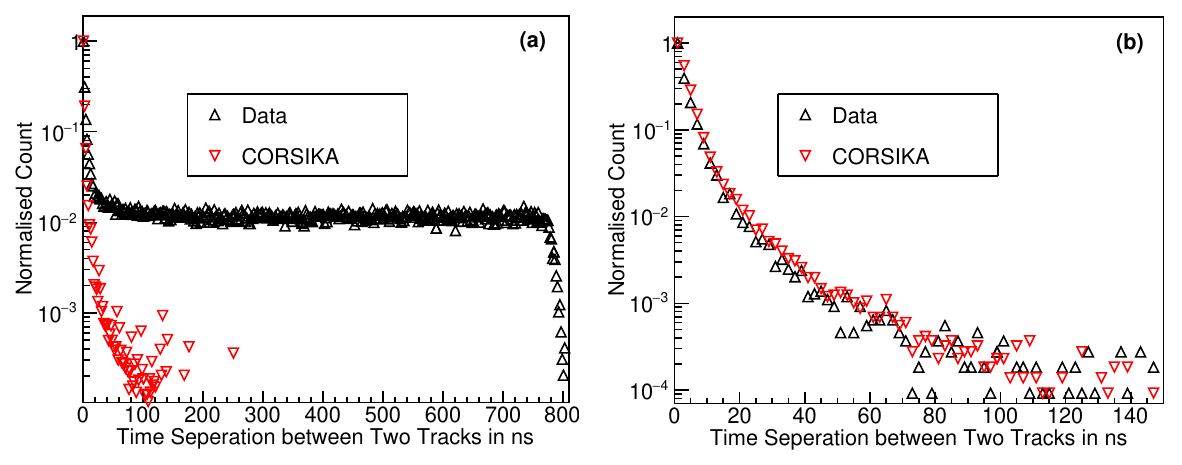} 
  \caption{Time separation of two tracks for (a) all events and (b) for events with 
           only parallel tracks.}
  \label{fig:time_sep}
\end{figure}

The events of interest for this analysis are the 
events with more than one reconstructed 3-dimensional track. The distribution of 
the time separation between each pair of tracks for both simulation and data are
shown in Figure~\ref{fig:time_sep}(a).
In the case of data, it can be observed that there is a significant number of
events where multiple particles are reaching the detector with large relative
time delay. The random coincidence of particles originating in the different
cosmic showers are the cause for these events. The random coincidence of particles
from different cosmic showers are absent in the simulation as only one shower is
simulated at a time in the CORSIKA.
So the following procedure is adopted to reject the random coincidences from
the events which have been initiated at the same showers.

\begin{figure}[h]
  \includegraphics[width=1.0\linewidth]{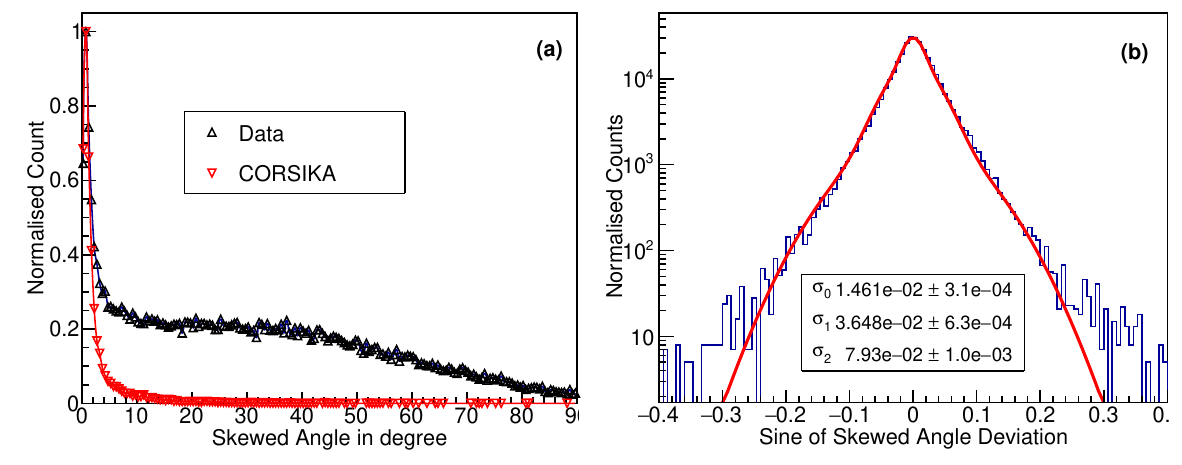} 
  \caption{(a) Skewed angle between two tracks originating outside of the detector, 
    (b) Skewed angle difference between generated and reconstructed tracks fitted
    with triple-Gaussian function.}
  \label{fig:skewed_angle}
\end{figure}

In the simulation, it is observed that the particles originating from the same
shower are detected in the RPC stack as parallel tracks. This can be verified by 
calculating the skewed angle between each pair of tracks reconstructed in an
event. The value of skewed angle is ideally supposed to be 
zero in case of parallel tracks, but due to finite size of the strip width and 
multiple scattering it has finite width and tails. The distribution of the skewed 
angle between each pair of tracks reconstructed in an event from both simulation
and data is shown in Figure~\ref{fig:skewed_angle}(a).
Now, only the parallel tracks are of importance in this study because of their
same origin. In order to define the parallel tracks reconstructed in this detector
setup, good understanding about the resolution of skewed angle is necessary.
To understand the application of the skewed angle, events with multiple
particles are simulated in the GEANT4. The skewed angle $\left(s_{gen}\right)$
between a generated pair of tracks is calculated using their generated directions.
The skewed angle $\left(s_{reco}\right)$ between the same pair of tracks is also
estimated from the track reconstruction. The distribution of the sine of the
difference of the skewed angle between the generated particles and the skewed angle
between the reconstructed tracks, defined as $\sin\left(s_{reco}-s_{gen}\right)$, is
shown in Figure~\ref{fig:skewed_angle}(b). This distribution is fitted with a
triple-Gaussian function. The three components of these angular resolutions
($\sigma_{0}$, $\sigma_{1}$, $\sigma_{2}$) represent the cases, where (0) no multiple
scattering happened for the pair of tracks, (1) one of the tracks has gone through
multiple scattering and (2) both the tracks have gone through multiple scatterings
in the detector medium or in the roof of the housing building, respectively.

Based on these observations, a pair of tracks with a skewed angle less than 
$2.5^{\circ} \left(\approx 3\sigma_{0}\right)$ are considered as parallel to each
other. All the pairs of tracks present in a reconstructed event has to comply with
this selection criteria. Thus, in
the current study, only the parallel tracks are considered to select the tracks 
generated from the particles originating from the same cosmic ray shower. The time 
difference between a pair of tracks for both simulated and observed data after 
the criteria of parallel track selection are shown in Figure~\ref{fig:time_sep}(b). 
It can be observed that the events from the random coincidences disappear after 
rejecting the events with non-parallel tracks.

The particles, originated in the interaction of the cosmic particles with the
detector medium or the roof are also not parallel to the cosmic particles or
with each others as the particles share a common vertex.
The skewed angle cut rejects these events as well.

\section{Results and Discussions} \label{sec:result}
\begin{figure}[h]
  \centering
  \includegraphics[width=0.85\linewidth]{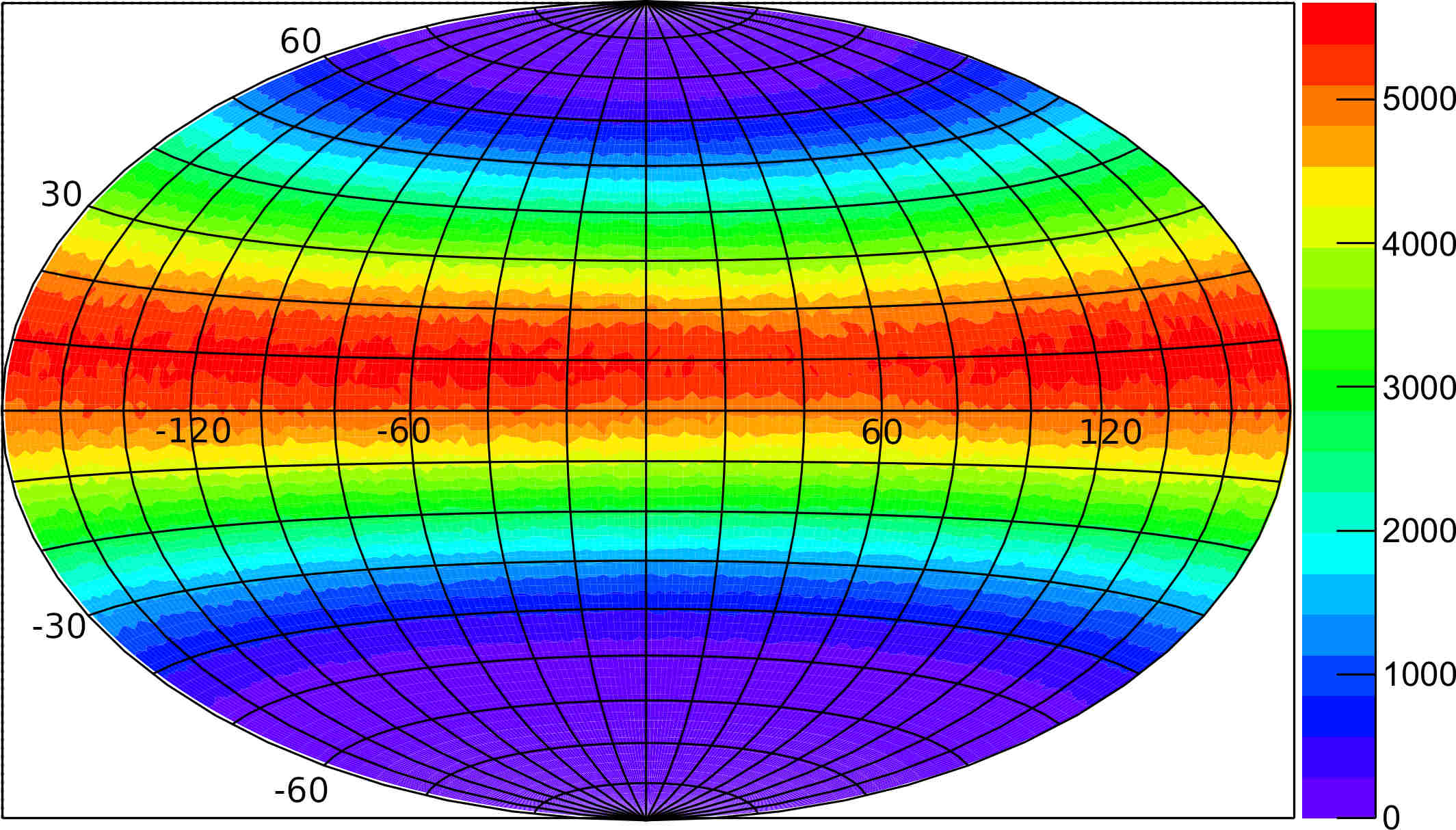} 
  \caption{Reconstructed direction of the tracks in the detector observed for the
    data recorded in the detector.}
  \label{fig:pinsk}
\end{figure}

In the present work, the event direction is presumed as a mean direction of all
individual muons in an event. In the analysis, the clustering of events towards any
specific region in the sky was not observed for the data recorded in the detector,
which can be confirmed by the Figure~\ref{fig:pinsk}.
Also, no significant modulation of the ratio of the events containing 2, 3 and 4
tracks with respect to single track events was observed during the period of
observation irrespective of the periodic changes in the reconstruction efficiency.
Hence, the assumption of uniform distribution of
cosmic ray directions which are used in the CORSIKA simulations are justified
by the absence of anisotropy in the data.

The total number of events with at least one reconstructed track in it is
approximately 206\,millions. The normalised fraction of the events containing
2, 3 and 4 tracks with respect to 
single track events are calculated to be $6.35\pm 0.05\times 10^{-5}$, 
$5.82\pm 0.53\times 10^{-7}$ and $1.94\pm 0.97\times 10^{-8}$, 
respectively from the cosmic ray data.

The normalised fraction of the events 
containing 2, 3 and 4 tracks are also calculated from the CORSIKA simulation for 
different types of cosmic primaries (H, He, C, O, Si, and Fe) and for different 
hadronic interaction models (QGSJET-II-04 and QGSJET01d), which are shown in
Table~\ref{tab:ratio1}. In order to compare the simulated results with data,
all the normalised fraction, calculated for different cosmic primaries are summed
with weights where the abundances in the primary
composition\cite{cosmic1,pdgspectra1} are used as the weights. The comparison of
data and the combined predictions are given in Table~\ref{tab:ratio2}.

\begin{table}[btbp]
 \scalebox{0.635}{
  \begin{tabular}{|c|cccccc|} \hline
   No of   &  H      & He       & C     & O    & Si      & Fe   \\  \cline{2-7}
   Tracks  & \multicolumn{6}{c|}{QGSJET-II-04}                 \\ \hline
          2      & $2.19\pm 0.12\times 10^{-5}$  & $4.71\pm 0.19\times 10^{-5}$ & $1.21\pm 0.02\times 10^{-4}$ & $1.61\pm 0.02\times 10^{-4}$ & $2.42\pm 0.02\times 10^{-4}$ & $4.58\pm 0.03\times 10^{-4}$ \\
          3      & $1.02\pm 0.12\times 10^{-7}$ & $3.04\pm 0.17\times 10^{-7}$ & $1.78\pm 0.05\times 10^{-6}$  & $3.11\pm 0.06\times 10^{-6}$ & $5.57\pm 0.08\times 10^{-6}$ & $1.61\pm 0.02\times 10^{-5}$ \\
          4      & $1.61\pm 0.65\times 10^{-9}$ & $8.80\pm 2.46\times 10^{-9}$ & $5.83\pm 0.47\times 10^{-8}$  & $1.12\pm 0.07\times 10^{-7}$ & $2.35\pm 0.11\times 10^{-7}$ & $1.02\pm 0.03\times 10^{-6}$ \\
\hline
                 & \multicolumn{6}{c|}{QGSJET01d}                     \\ \hline
          2      & $2.14\pm 0.12\times 10^{-5}$ & $4.74\pm 0.13\times 10^{-5}$ & $1.19\pm 0.02\times 10^{-4}$  & $1.52\pm 0.02\times 10^{-4}$ & $2.50\pm 0.02\times 10^{-4}$ & $4.56\pm 0.03\times 10^{-4}$ \\
          3      & $9.13\pm 1.22\times 10^{-8}$ & $3.91\pm 0.18\times 10^{-7}$ & $1.90\pm 0.04\times 10^{-6}$  & $3.14\pm 0.08\times 10^{-6}$ & $6.19\pm 0.07\times 10^{-6}$ & $1.65\pm 0.02\times 10^{-5}$ \\
          4      & $0.75\pm 0.38\times 10^{-9}$& $6.48\pm 1.38\times 10^{-9}$ & $6.00\pm 0.43\times 10^{-8}$  & $1.07\pm 0.07\times 10^{-7}$ & $3.39\pm 0.11\times 10^{-7}$ & $1.16\pm 0.03\times 10^{-6}$ \\ \hline
\end{tabular}}
 \caption{Fraction of track with 2, 3 and 4 tracks obtained from Simulation 
   for different primaries (H, He, C, O, Si and Fe) and different physics packages 
   (QGSJET-II-04 and QGSJET01d).}\label{tab:ratio1}
\end{table}

\begin{table}[btbp]
  \centering
\begin{tabular}{|c|ccc|} \hline
No of Tracks &  Data &   QGSJET-II-04     &  QGSJET01d \\ \hline
    2         & $6.35\pm 0.05\times 10^{-5}$ & $2.35\pm 0.13\times 10^{-5}$ & $2.37\pm 0.12\times 10^{-5}$ \\
    3         & $5.82\pm 0.53\times 10^{-7}$ & $1.12\pm 0.13\times 10^{-7}$ & $1.23\pm 0.13\times 10^{-7}$ \\
    4         & $1.94\pm 0.97\times 10^{-8}$ & $3.21\pm 0.87\times 10^{-9}$ & $2.43\pm 0.50\times 10^{-9}$ \\ \hline
\end{tabular}
\caption{Comparison of track fraction with 2, 3 and 4 tracks obtained from Data
  and simulation.}\label{tab:ratio2}
\end{table}

If the abundances of elements in the primary cosmic ray spectrum as observed in 
\cite{cosmic1,pdgspectra1} are used to form the final result from simulation, the 
normalised track fractions of the events containing 2, 3 and 4 tracks are within
one order of magnitude less compared to the data.
Systematic error due to uncertainties of roof thickness, material in the detector 
setup, strip multiplicity, noise, efficiencies and the physiscs models used in GEANT4
are much smaller than the observed discrepancy between data and MC prediction.
These results clearly demonstrate that there is a discrepancy between the
observed data and predictions from the cosmic ray particle spectrum, the CORSIKA
and finally the GEANT4 simulation.

A few other experiments (KGF\cite{kgf1}, ALICE\cite{alice1}, MACRO\cite{macro1}, 
DELPHI\cite{delphi1}, ALEPH\cite{aleph1}, KASCADE-Grande\cite{kascade1}, etc.) have 
also studied the multi-muon tracks in cosmic events. Except for the KASCADE-Grande, 
all other experiments were performed under the ground. The underground experiments 
have observed events with large multiplicities because of the large size of the 
detectors and the overburden of rock and soil, which are blocking showers with 
low energy. The multiplicity of the cosmic ray particles observed in a detector
is highly dependent on the dimensions, aperture, energy threshold and detector's
location. Hence, it is difficult to compare the results of the aforesaid
experiments quantitatively with the small-scale detectors setup in this current
study.
But all the studies based on the aforesaid experiments have indicated a similar 
discrepancy between the CORSIKA spectra and the observed data. The KASCADE-Grande 
experiment has also concluded that the attenuation length of muons in the 
atmosphere from the simulation is smaller than estimation from the observed 
data\cite{kascade1}.

Although, the bulk of primary particles interacts at center-of-mass energies far
below 10\,GeV \cite{corsika763}, it is observed in the simulation that a significant
amount of primaries which are responsible for higher particle multiplicities in
the present setup, are in the range beyond the scope of the present collider
experiments.
The major problem of the EAS simulation programs is the extrapolation of the 
hadronic interactions in the high energy range which is not covered by the 
experimental data. The limitation of the experiments to measure the hadronic 
interactions at this high energy is mainly due limitation of the design of high
energy $p\bar{p}$-colliders \cite{corsika763}. In the present $p\bar{p}$-colliders,
the forward direction is not accessible.
The secondary particles which are of the higher 
importance in the development of EAS programs are undetected in the beam pipe of 
the colliders. The largest energy fraction of each $p\bar{p}$-collision is carried 
away by these particles. The maximum attainable energy in these colliders is much 
lower than those found in cosmic rays. Therefore, the extrapolations based on 
theoretical models are mainly used by all the EAS programs.
While, the energy of the cosmic rays contributing for the single track events in
the current detector setups are well within the boundary of the current physics models
(upto $\sim$1\,TeV), a significant amount of the interactions contributing for the
higher multiplicities are beyond this energy and are not supported by the experimental
data \cite{corsika763}.

\section{Conclusion}
In the period between August 23, 2017, to September 8, 2017, approximately
250 millions events were recorded. There is a discrepancy in results
predicted by the EAS simulation program which can be observed in the 
comparison of the track multiplicity between the data and Monte-Carlo.
The results of the current study reflect that the current physics models of 
interactions at the earth atmosphere are unable to reproduce the air showers 
accurately. The earlier measurements of muon multiplicity along with the present 
result can be used to improve the parameters of the hadronic model at high 
energies and/or cosmic ray spectral index along with the composition of the
primary cosmic rays.

\end{document}